\documentclass{aa}
\usepackage[utf8]{inputenc}

\usepackage{graphics,epsfig,txfonts}
\usepackage[section]{placeins}
\usepackage{natbib}
\bibpunct{(}{)}{;}{a}{}{,}

\usepackage{color}

\begin{document}
 
\title{eROSITA cluster cosmology forecasts: cluster temperature substructure bias}

\author{      F.~Hofmann\inst{1}
     \and     J.S.~Sanders\inst{1}
     \and     N.~Clerc\inst{1,2,3}
     \and     K.~Nandra\inst{1}
     \and     J.~Ridl~\inst{1}
     \and     K.~Dennerl~\inst{1}
     \and     M.~Ramos-Ceja~\inst{4}
     \and     A.~Finoguenov~\inst{1,5}
     \and     T.H.~Reiprich~\inst{4}
       }

\titlerunning{eROSITA simulated cluster observations}
\authorrunning{Hofmann et al.}

\institute{Max-Planck-Institut f\"ur extraterrestrische Physik, Giessenbachstra{\ss}e, 85748 Garching, Germany
\and CNRS; IRAP; 9 Av. colonel Roche, BP 44346, F-31028 Toulouse cedex 4, France
\and Université de Toulouse; UPS-OMP; IRAP; Toulouse, France
\and Argelander-Institut f\"ur Astronomie, University of Bonn, Auf dem H\"ugel 71, 53121, Bonn, Germany
\and Department of Physics, University of Helsinki, PO Box 64, 00014, Helsinki, Finland
}


\abstract{The eROSITA mission will provide the largest sample of galaxy clusters detected in X-ray to date (one hundred thousand expected). This sample will be used to constrain cosmological models by measuring cluster masses. An important mass proxy is the electron temperature of the hot plasma detected in X-rays.}
	 {We want to understand the detection properties and possible bias in temperatures due to unresolved substructures in the cluster halos.}
	 {We simulated a large number of galaxy cluster spectra with known temperature substructures and compared the results from analysing eROSITA simulated observations to earlier results from \emph{Chandra}.}
	 {We were able to constrain a bias in cluster temperatures and its impact on cluster masses as well as cosmological parameters derived from the survey. We found temperatures in the eROSITA survey to be biased low by about five per cent due to unresolved temperature substructures (compared to emission-weighted average temperatures from the \emph{Chandra} maps). This bias would have a significant impact on the eROSITA cosmology constraints if not accounted for in the calibration.}
	 {We isolated the bias effect that substructures in galaxy clusters have on temperature measurements and their impact on derived cosmological parameters in the eROSITA cluster survey.}

\keywords{Galaxies: clusters -- X-rays: galaxies: clusters}

\maketitle

\section{Introduction}
\label{sec:introduction}

Clusters of galaxies reveal the large-scale structure of the universe and allow us to observe astrophysical processes on large scales. They are among the largest gravitationally bound structures observable in the universe and one of the most sensitive methods for detecting them is by the X-ray radiation of the hot intra-cluster-medium (ICM). The importance of cluster observations for cosmological studies has been proven by the first X-ray all-sky survey with the ROSAT satellite \citep[][]{1982AdSpR...2..241T}. This survey allowed to observe a population of $\rm{\sim900}$ clusters across the sky \citep[e.g.][]{2004A&A...425..367B,2002ApJ...580..774E,2000ApJS..129..435B}. The mass distribution of this population can be used to constrain cosmological models. Before the start of the ROSAT mission, there were detailed simulations of cluster observations \citep[][]{1991ExA.....1..365C} in order to estimate the total number of expected cluster detections during the survey and to prepare for the data analysis once the real data was available.

Since the end of the ROSAT mission, many of the clusters originally detected with ROSAT have been observed deeper and at higher spectral and spatial resolution with the \emph{Chandra} and \emph{XMM-Newton} X-ray observatories. However these telescopes only observed individual clusters and in a small fraction of the complete sky.

The all-sky X-ray survey mission, the extended Roentgen Survey with an Imaging Telescope Array (eROSITA) on the Spectrum-Röntgen-Gamma (SRG) satellite \citep[][]{2012arXiv1209.3114M,2010SPIE.7732E..0UP} will perform a $\gtrsim 20$ times more sensitive survey in the $\rm{0.5-2.0~keV}$ ROSAT X-ray band and the first truly imaging survey for energies from $\rm{2-10~keV}$. The instrument consists of seven X-ray telescopes with separate detector arrays.

The main science goal of the mission is the detection of the largest sample of galaxy clusters ($\rm{\sim 10^5}$) out to a redshift of $z \gtrsim \mathrm{1}$. This sample will deliver strong constraints on cosmological models and their parameters, especially dark energy. It is very important to understand the characteristics of the clusters, which will be detected with eROSITA. 
First simulations of clusters in the eROSITA survey were made to derive estimates on the number of clusters and the general reliability of cluster temperatures \citep[e.g.][]{2012MNRAS.422...44P,2014A&A...567A..65B}. \citet{2016A&A...585A.130H} analysed deep \emph{Chandra} observations and derived emission models for 33 clusters with very high spatial resolution of temperature structures. We used these cluster models to simulate eROSITA observations and to identify bias due to unresolved substructures in cluster temperatures in the eROSITA survey caused by the lower spatial resolution of the eROSITA instrument compared to \emph{Chandra}.

Unless stated otherwise we used a standard $\rm{\Lambda CDM}$ cosmology with $\rm{H_0=71~km~s^{-1}~Mpc^{-1}}$, $\rm{\Omega_M=0.27}$ and $\rm{\Omega_{\Lambda}=0.73}$ and relative solar abundances as given by \citet{1989GeCoA..53..197A}.

\section{Sample properties}
\label{sec:sample}

We created simulated eROSITA observations for a well defined and analysed sample of \emph{Chandra}-observed clusters of galaxies \citep[see sample by][]{2016A&A...585A.130H}. 
The clusters in the sample have halo masses ranging from $\rm{1\times10^{14}~M_{\odot}}$ to $\rm{2\times10^{15}~M_{\odot}}$ (within the overdensity-radius $r_\mathrm{500}$,  where the average density of the cluster is 500 times the critical density of the universe at the cluster redshift). The luminosity range is $\rm{(2-63)x10^{44}~erg/s}$ ($\rm{0.1-2.4~keV}$ X-ray luminosity), and the redshift ranges from 0.025 to 0.45 with a median of 0.15.
The sample contains a large variety of cluster structures and therefore is representative of massive evolved clusters in the universe.

The eROSITA survey will have an average half energy width (HEW) of $\rm{\sim28~arcsec}$ at 1\,keV (measure for the average extent of a point source in the survey), the energy range will be about $\rm{0.2-8.0~keV}$ (see Fig. \ref{fig:arfnew}), and the spectral resolution will be $\rm{\sim138~eV}$ at 6\,keV. The main survey will last four years and is expected to provide a catalog of $\rm{\sim 10^5}$ galaxy clusters with a median redshift of 0.35 \citep[][]{2012MNRAS.422...44P} and a median cluster mass of $\rm{\sim 10^{14}~M_{\odot}}$. 
For a subset of eROSITA clusters ($\rm{\sim 1500}$) the temperature $T_\mathrm{X}$ can be measured with an uncertainty of less than 10 per cent \citep[see estimates by][]{2014A&A...567A..65B}.
This subsample will have higher median mass and lower median redshift than the total sample, justifying our comparison to the \emph{Chandra} sample for the $T_\mathrm{X}$ bias study. Even for such a reduced sample the temperature bias of five per cent found in this study (see Sect.\ref{sec:tbias}) would still create a significant offset in derived cosmological parameters. Stacked spectra and binning of clusters will allow temperature measurements for an even larger sample in the final four-year survey of eROSITA.

\section{Simulated spectra with XSPEC}
\label{sec:simulations_spec}

We used the latest response files and exposure estimates available to the eROSITA consortium (state of the art on July 14, 2016) for the simulations of cluster spectra in the four year eROSITA survey.

\begin{figure}
  \centering
  \resizebox{0.99\hsize}{!}{\includegraphics[angle=0,trim=1cm 13.5cm 1cm 4cm,clip=true]{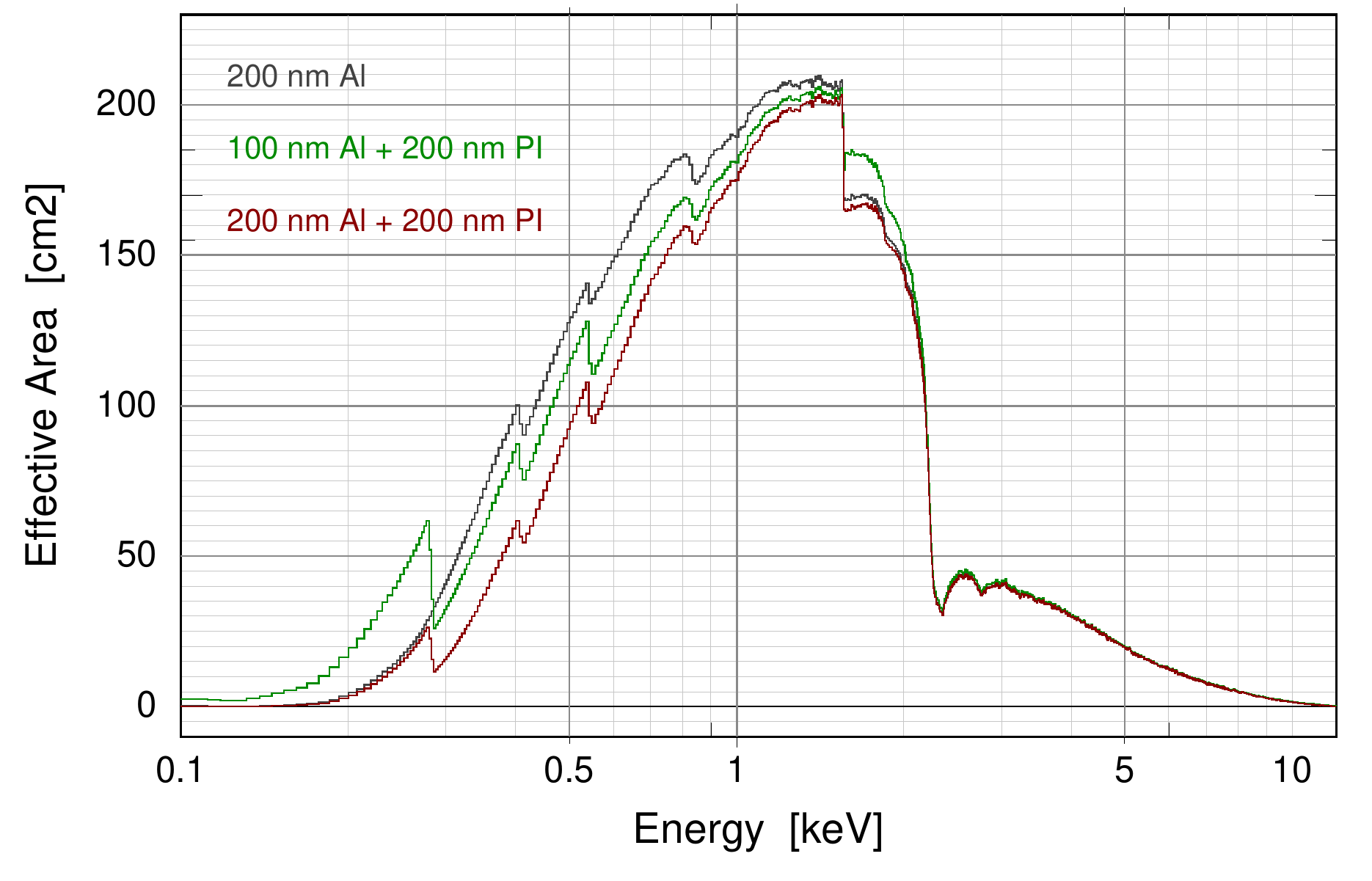}}
  \caption[Latest eROSITA ARF calibration]{Latest measurements of eROSITA ARF for different filter combinations (preliminary). The plot shows the effective area over energy for one of seven telescopes taking into account the mirror, filter transmission and CCD quantum efficiency. The response is averaged over the FoV of the telescope.}
  \label{fig:arfnew}
\end{figure}

We simulated the spectra of the clusters from the best fit values of the \emph{Chandra} cluster sample maps in each spectral-spatial region \citep[][]{2016A&A...585A.130H}. 
The cluster maps were created with the contour binning technique of \citet{2006MNRAS.371..829S}.

To validate the simulation and fitting methods we made two identical simulations which were processed with the exact same analysis procedure.
Of the two simulations, one contains the cluster substructure as measured from the deep \emph{Chandra} sample, and the other an isothermal cluster model with the median temperature of the cluster emission of the substructure case. This median temperature of the cluster maps corresponds to an emission weighted temperature of the cluster (with about one per cent scatter), because the spatial bins of the cluster maps were created with a constant signal-to-noise level, meaning the number of bins in a region is proportional to its X-ray brightness.

Each spectrum was simulated using \texttt{XSPEC} (version 12.9.0o) \texttt{fakeit} which creates a simulated instrument spectrum of a emission model applying an instrument response and Poisson noise on the counts in each spectral channel of the detector. 
The eROSITA spectra have 1024 energy channels with a width of $\rm{\sim50~eV}$. 
Each spectrum was created with an absorbed \texttt{apec} model which is defined by normalisation, temperature, foreground absorption, redshift, and metallicity. Normalisation, temperature, and foreground absorption (by neutral hydrogen) were taken from the \emph{Chandra} cluster sample maps \citep[][]{2016A&A...585A.130H}. The foreground absorption by neutral hydrogen column density between observer and source was modelled by the \texttt{phabs} model included in \texttt{XSPEC}. The redshift was fixed according to the simulated redshift bin of the simulation. The metallicity was fixed at the average sample value of $Z/Z_{\sun} = \mathrm{0.3}$. The normalisation of a spectrum is directly proportional to the count flux of the source and can be converted to an energy flux if the source spectrum is known.  
The normalisation of the spectrum was scaled according to the change in luminosity distance between the \emph{Chandra} measurement and the simulated eROSITA observation.
The simulated spectra do not include background effects.
The exposure of the simulated spectra is 2\,ks which is the average exposure of the most recent survey strategy for the all-sky survey after four years.

For the response files we assumed seven identical telescopes with a 200\,nm on-chip aluminium (Al) filter (see top curve in Fig.\ref{fig:arfnew}). Fig.\ref{fig:arfnew} shows the effective X-ray collecting area (ARF file) for one telescope. For seven identical telescopes the average effective area between $\rm{1-2~keV}$ will be about $\rm{1400~cm^{2}}$. 

The second calibration file needed for the simulation is the RMF, which is a two
dimensional matrix describing the probability with which an incoming photon of a
specific energy is measured in a certain energy channel of the detector.
We used an RMF averaged over all split events (accounting for measured split event fractions) for the 200\,nm on-chip Al filter case. Response files have been measured with the flight hardware and put together into a format readable by the \texttt{XSPEC} fitting package. These have been provided internally to the German eROSITA consortium in May 2015.
Every cluster was simulated 100 times and at 5 different redshifts (0.1, 0.2, 0.4, 0.8, 1.6).

\section{X-ray mass-proxy bias}
\label{sec:proxybias}

We investigated a possible bias in measured cluster temperature $T_\mathrm{X}$ and flux $F_\mathrm{X}$ induced by the substructures of the ICM temperature. We applied a Bayesian parameter estimation technique \citep[BXA, see][]{2014A&A...564A.125B} to all simulated spectra to obtain a distribution of median values for the X-ray cluster temperature $T_\mathrm{X}$ and flux $F_\mathrm{X}$ in the $\rm{0.3-6.0\,keV}$ band. 

The spectra between $\rm{0.3-6.0~keV}$ were loaded into \texttt{XSPEC}, the background is set to zero, and the a \texttt{apec} x \texttt{phabs} model is initialized. The priors for the BXA fitting procedure are uniform distributions between fixed limits. The limits for $T_\mathrm{X}$ were set to ${\rm1.0-20.0~keV}$.
With these priors, BXA is run to obtain the distribution of values for norm and $T_\mathrm{X}$. 
For each BXA iteration the flux in the $\rm{0.3-6.0~keV}$ band is calculated.
From the BXA distributions we obtained the median value, upper-, and lower-bound as percentiles of the distributions of $T_\mathrm{X}$ and $F_\mathrm{X}$ (15, 50, and 85 per cent as lower, median, and upper values). This corresponds to best fit and $\rm{\sim1\sigma}$ range for a Gaussian distribution. 

We simulated every cluster spectrum 100 times to estimate the influence of statistical fluctuations on our results and thus obtained 100 median values for each parameter. For the further analysis we plot the percentiles (15, 50, and 85 per cent) of the distribution of the 100 median values. This is a good approximation of what will be measured for a certain type of cluster in the eROSITA survey.

\subsection{Bias in temperature $T_\mathrm{X}$}
\label{sec:tbias}

For estimating eROSITA temperature bias due to temperature substructure in the cluster ICM we used the output of the \texttt{XSPEC} simulations described above. 
At five redshifts (0.1, 0.2, 0.4, 0.8, and 1.6) we simulated the sample of 33 clusters both with real substructure (sub) and as isothermal (iso) clusters. Each simulation was done 100 times and we extracted the distribution of the median values in temperature. We calculated the significance of offset between the distribution of the real and isothermal cluster cases as,

\begin{equation}
 \mathrm{Bias}(T) = \frac{T_\mathrm{sub}-T_\mathrm{iso}}{T_\mathrm{average}}
\end{equation}

where $T_\mathrm{sub}$ is the temperature measured in the simulation with substructure, $T_\mathrm{iso}$ is the temperature measured in the isothermal case, and $T_\mathrm{average}$ is the average temperature of the two.

\begin{figure}
  \centering
  \resizebox{0.9\hsize}{!}{\includegraphics[angle=0,trim=0cm 0cm 0cm 0cm,clip=true]{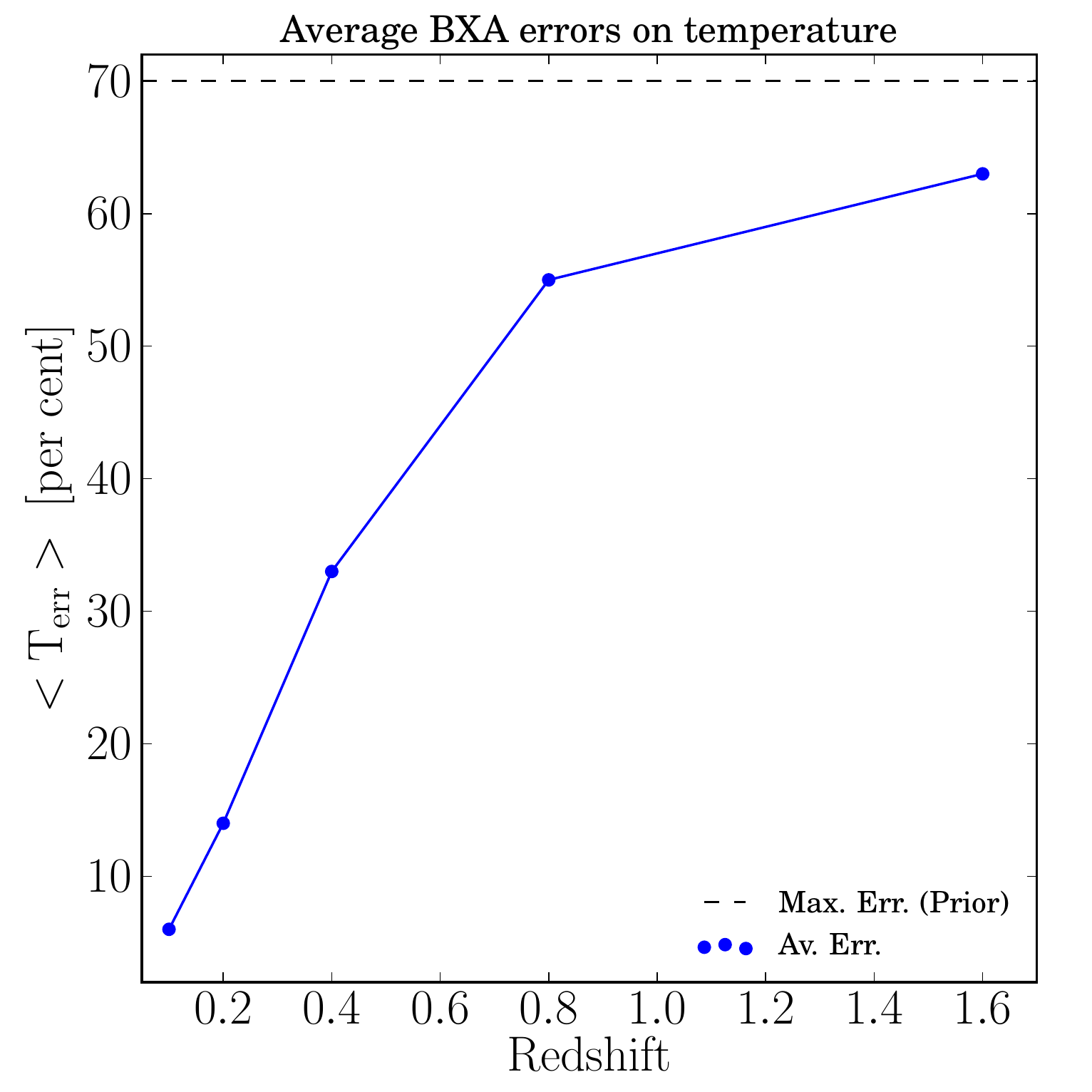}}
  \caption[Average statistical error on eROSITA temperature measurements]{Average error on eROSITA temperature measurements in individual clusters in the all-sky survey. The dotted line indicates the maximum error (temperature measurement not possible) using the BXA fitting approach.}
  \label{fig:terr_red}
\end{figure}

The sub-iso measurements allowed us to isolate the bias due to temperature substructure.
Fig.\ref{fig:terr_red} shows the evolution of the statistical measurement uncertainties increasing with redshift. We assumed a prior on temperature ($\rm{1-20~keV}$) and if the temperature can not be constrained the probability distribution becomes flat. Because best fit and uncertainties are extracted using percentiles of the distribution the best fit value then tends towards the middle of the prior at 10.5\,keV. The maximum uncertainty in this case is $\rm{\sim70~per~cent}$.

The analysis does not show significant difference ($\rm{\lesssim1\sigma}$) in the temperature bias between different substructure types. However there was a slight trend of less bias in more disturbed systems. This can be understood because the low temperature components of cool cores (CC) are over-weighted in the soft eROSITA X-ray spectra and more disturbed systems have generally higher temperatures where the effect is smaller \citep[see e.g.][]{2013SSRv..177..195R}.

\begin{figure}
  \centering
  \resizebox{0.9\hsize}{!}{\includegraphics[angle=0,trim=0cm 0cm 0cm 0cm,clip=true]{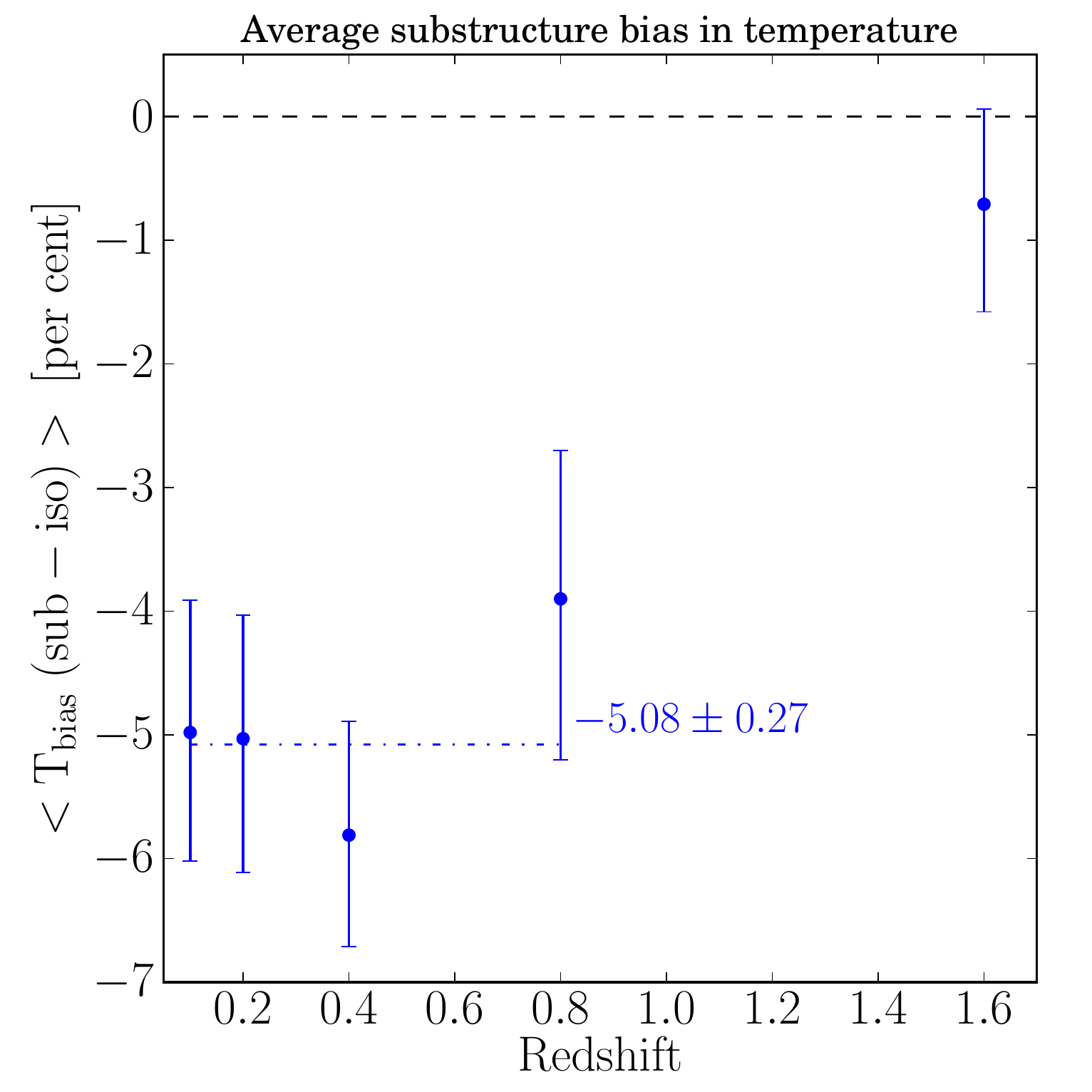}}
  \caption[Redshift dependent eROSITA temperature bias]{Temperature bias due to substructure at different redshifts. The annotation shows the average bias between redshift 0.1 and 0.8. Errorbars are $\rm{1\sigma}$ uncertainties from a bootstrap re-sampling technique.}
  \label{fig:bias_red}
\end{figure}

Fig.\ref{fig:bias_red} shows the average bias for the cluster sample at different redshifts. Because temperature uncertainties increase at higher redshift, we could not measure a significant offset between iso and sub at redshift $z \sim \mathrm{1.6}$. Most clusters will be below redshift 0.8 in the eROSITA survey so it will be the most important range for bias correction.

The measured bias from simulations is $\rm{-5.08\pm0.27}$ per cent in the redshift range $\mathrm{0.1} \leq z \leq \mathrm{0.8}$.

\subsection{Bias in flux $F_\mathrm{X}$}
\label{sec:fbias}

From the analysis of the simulated eROSITA spectra we obtained a distribution of best fit X-ray fluxes ($\rm{0.3-6.0~keV}$ energy band).

\begin{figure}
  \centering
  \resizebox{0.9\hsize}{!}{\includegraphics[angle=0,trim=0cm 0cm 0cm 0cm,clip=true]{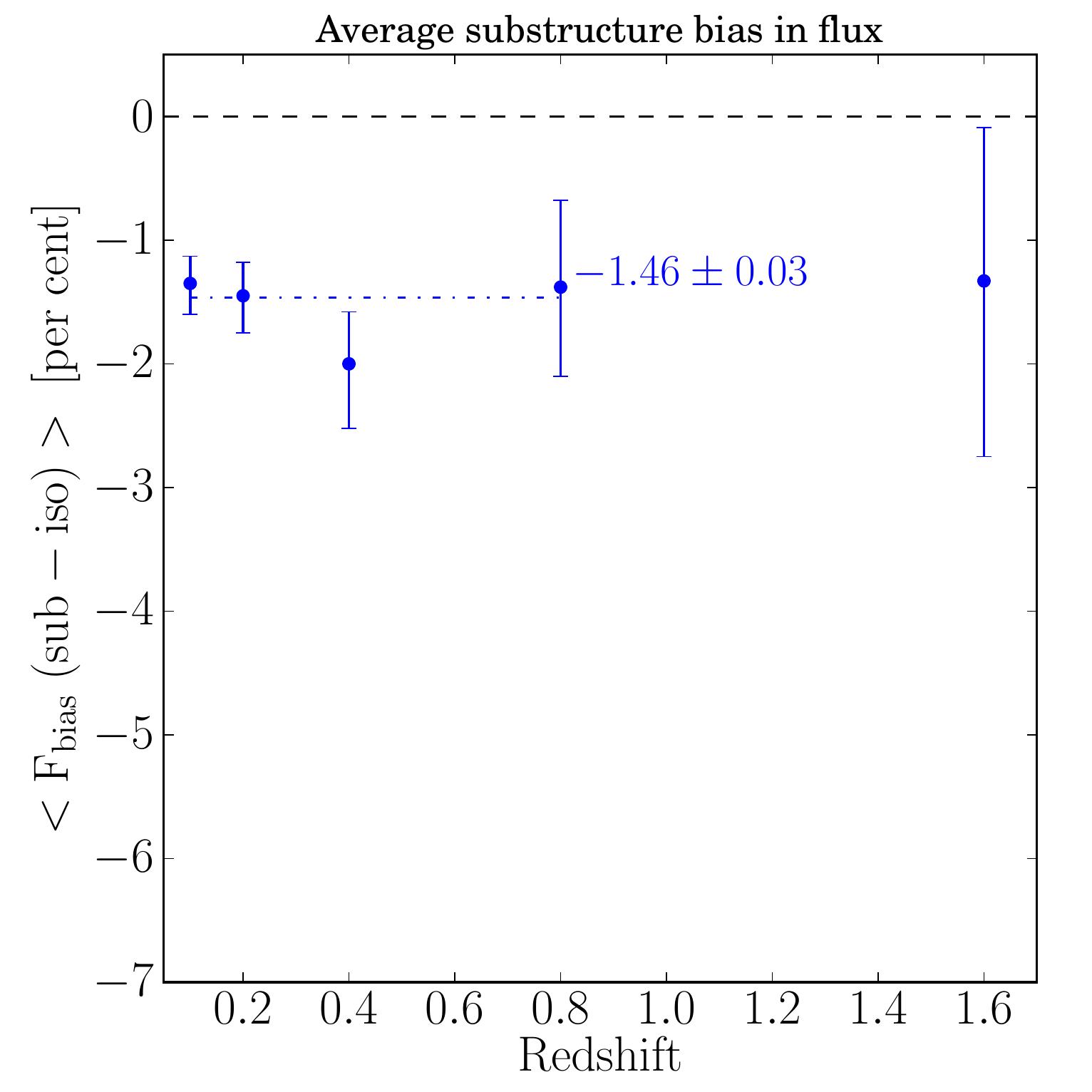}}
  \caption[Redshift dependent eROSITA flux bias]{Flux bias due to temperature substructure at different redshifts. The annotation shows the average bias between redshift 0.1 and 0.8. Errorbars are $\rm{1\sigma}$ uncertainties from a bootstrap re-sampling technique.}
  \label{fig:fbias_red}
\end{figure}

Fig.\ref{fig:fbias_red} shows that the bias in flux is lower than in temperature but there is also a significant offset due to temperature substructures. This can be due to correlation of the temperature and normalisation of a fit in the \texttt{apec} model. This correlation is clearly present in the parameter chains created by the BXA fitting procedure.
If the measured temperature is lower also the flux will be underestimated by a certain amount. The bias in flux is about 30 per cent of the bias in temperature. 

The measured flux bias is $\rm{-1.46\pm0.03}$ per cent in the redshift range $\mathrm{0.1} \leq z \leq \mathrm{0.8}$. Bias in flux translates directly to bias in X-ray luminosity $\rm{L_X}$ which is an important mass-proxy for galaxy clusters \citep[L-M scaling relations, see e.g.][]{2009A&A...498..361P,2011A&A...526A.105Z,2013MNRAS.435.1265E}. However the redshift of the source has to be known to determine intrinsic luminosity from observed flux. 

\subsection{Substructure dependence}
\label{sec:substr_types}

There is a large range of different substructure types in the \emph{Chandra} cluster sample. These types are mainly characterised by how strongly disturbed the ICM of a clusters is. There can be strong AGN feedback where jets from the AGN in the central galaxy heat the surrounding ICM and perturb the hydrostatic equilibrium of the system causing large temperature asymmetries. Mergers with subhalos can cause strong perturbations and shocks or sloshing motions.

We divided the clusters into three different substructure types by eye using the high resolution maps from \emph{Chandra} observations \citep[][]{2016A&A...585A.130H}: CC, disturbed, and double peaked clusters (see Tab. \ref{tab:bias}).
These substructure types provide an estimate of how perturbed the hot ICM halo is but the transitions between the three types are smooth. 
We found no significant difference in the bias between the three substructure types but a slight trend of less bias in more disturbed systems (about $\rm{1\sigma}$ significance).
This could be caused by higher average temperatures and thus a smaller bias effect towards lower energies.
The effect was strongest for CC systems with lower-temperature components at the center.

\begin{table}
\caption[]{Temperature bias values for different substructure types.}
\begin{center}
\begin{tabular}{lr}
\hline\hline\noalign{\smallskip}
  \multicolumn{1}{l}{Cluster type \tablefootmark{a}} &
  \multicolumn{1}{l}{$\mathrm{Bias}(T)~[\mathrm{per~cent}]$} \\
\noalign{\smallskip}\hline\noalign{\smallskip}
  CC & $\mathrm{-6.0\pm1.6}$ \\
  Disturbed & $\mathrm{-4.9\pm1.5}$\\
  Double peaked & $\mathrm{-2.8\pm1.9}$\\
\noalign{\smallskip}\hline
\end{tabular}
\tablefoot{
\tablefoottext{a}{Clusters per type \citep[abbreviations see][]{2016A&A...585A.130H} - CC: a1795, a1835, ms1455, a1413, ms0735, a2204, cygnusa, a907, 2a0335, a2597, a1650, a2199, hydraa, 3c348, and a2052; disturbed: a1995, rxj1347, zw3146, a1689, a401, pks0745, a2034, a3667, a496, sersic159, and a2390; double peaked: 1e0657, a665, a520, a2146, a521, a1775, and a2744.}
}
\end{center}
\label{tab:bias}
\end{table}

\subsection{Caveats}
\label{sec:bias_cav}

The purpose of this study was to isolate the influence of substructure and show a first order impact on cluster cosmology constraints for eROSITA. To obtain a more realistic and complete bias estimate for the eROSITA survey several other (possibly stronger) factors have to be considered, like the selection function of clusters in the eROSITA survey. For this purpose there are additional papers in preparation (e.g. Clerc et al., in prep.).

\begin{figure}[ht]
  \centering
  \resizebox{0.99\hsize}{!}{\includegraphics[angle=0,trim=0cm 0cm 0cm 0cm,clip=true]{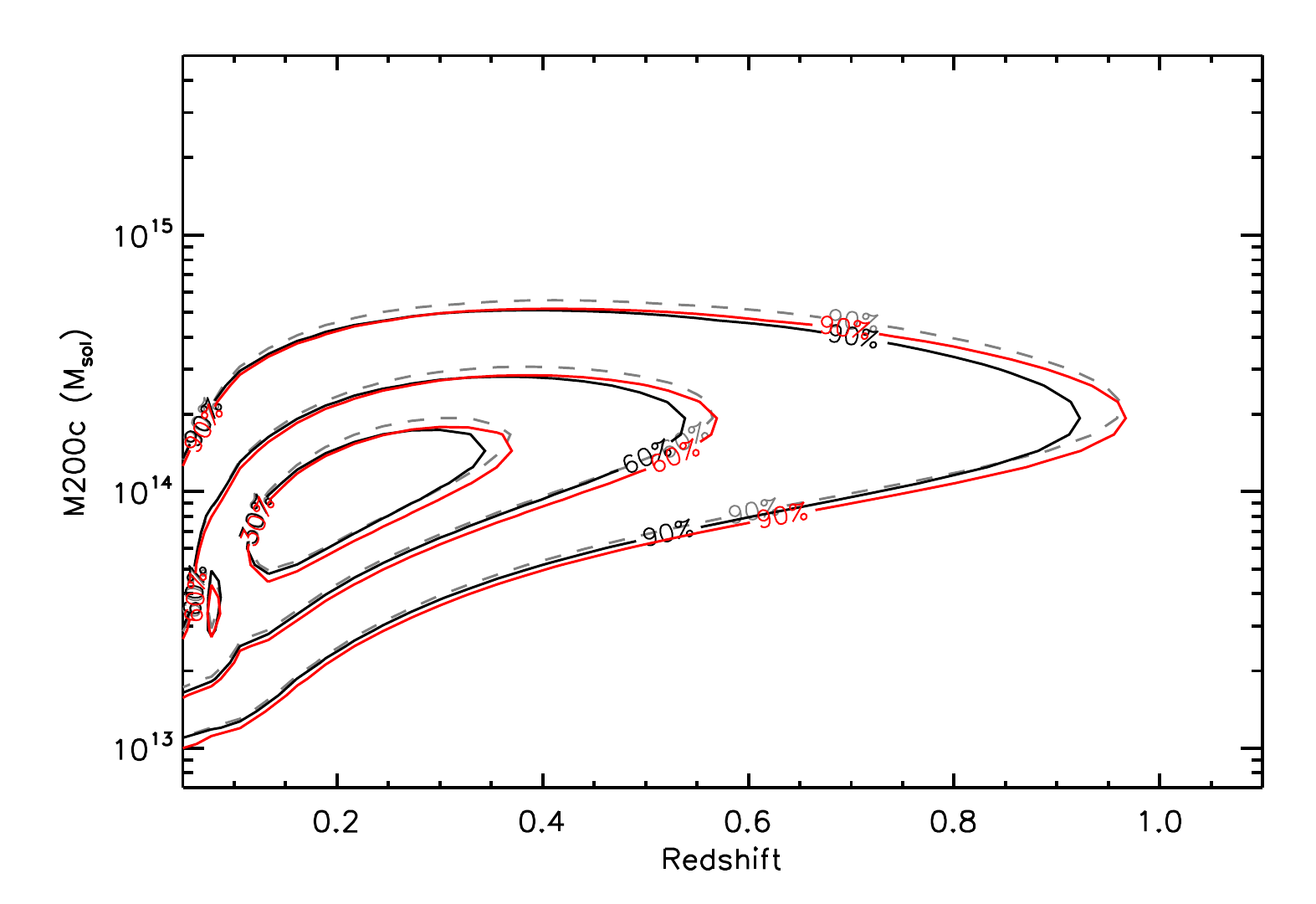}}
  \caption[Expected eROSITA mass function bias due to substructure]{Simulation of observed eROSITA cluster density in the redshift - mass ($\rm{M200c}$) plane. The total number of clusters is $\rm{\sim10^5}$. The contours include the denoted fraction of total observed clusters in the four year all-sky survey. The grey-dotted (higher in mass) contours show the expected observed distribution accounting for the selection function of the eROSITA survey. The red (lower in mass) contours show the distribution as it will be measured with a five per cent low bias in X-ray halo temperatures. The solid black contours indicate best-fit mass function (see below).}
  \label{fig:massfuntion}
\end{figure}

Additional bias can arise from effects which have not been accounted for in the \texttt{XSPEC} simulations. These include background in the observations and contamination of cluster spectra by AGN as well as uncertainty in the redshift of the clusters.  \citet{2007A&A...472...21L} found the temperature bias to increase when background is added to observations. Missing part of the cluster emission due to detection efficiency can cause additional bias. The measured cluster masses from fitting models to the spectra can also be biased by inaccuracies in the models or assumptions made which do not apply to the investigated system (e.g. non-thermal pressure support).

The \texttt{XSPEC} simulations only provide blended spectra without background and can thus not be used for core-excising tests or other spatially resolved temperature analysis.

The luminosity of a cluster will contain additional bias from the redshift measurements. It will depend on the quality and quantity of the available cluster redshifts whether $\rm{L_X}$ or $T_\mathrm{X}$ will be the better mass-proxy in the eROSITA survey. The current accuracy of photometric cluster redshifts is $\mathrm{0.01} \lesssim \Delta z / (\mathrm{1}+z) \lesssim \mathrm{0.02}$ \citep[see e.g. recent work by][and Klein et al., submitted]{2016ApJS..224....1R,2017arXiv170204314R}.

In the special case of bias due to temperature substructures it does not seem to be necessary to correct for different bias depending on substructure types. From first tests it was not clear whether there is a reliable measurement for quantifying substructure in the eROSITA survey. This is mainly due to Poisson noise in the relatively shallow observations and a relatively large average survey PSF which will blur any substructure features in the cluster emission.

Because the bias does not have a strong dependence on the cluster type it is not critical how representative of the real cluster population of the universe the 33 cluster sample is. The sample used in this study covers redshifts from 0.05 to 0.45 and contains relatively high mass systems with a large variety of substructure types. Even if the fraction of different types is not perfectly representative we estimate the systematic uncertainty of the bias measurement to be less than one per cent.

In the simulations we assumed that the cluster sample is the same in each redshift bin. This means that evolution in the cluster properties over time can not be probed. For example evolution in mass profiles \citep[e.g][]{2008MNRAS.386.1045A,2010AIPC.1241..192B} or the ratio of CC clusters to merging systems in the course of hierarchical halo formation \citep[see e.g. from cosmological simulations][]{1997ApJ...490..493N}.

The final filter configuration of eROSITA will be 5 x 200\,nm Al (on-chip) and 2 x 100\,nm Al, placed behind 2-7 x 200\,nm Polymid (PI) filters. This will slightly change the instrument response files compared to our simulations where we assumed 7 x 200\,nm Al, but will not influence the temperature bias results significantly.

\section{Mass function and cosmology bias}
\label{sec:massf_cosmo}

The temperature of the hot ICM measured from X-ray spectra is an important proxy for the mass of the observed cluster \citep[for a recent review on different mass-proxies, see e.g.][]{2012ARA&A..50..353K}. Temperature can only be measured accurately in high S/N spectra. The X-ray flux from a cluster can be measured accurately also in observations with lower S/N and can be used to estimate the mass \citep[see e.g. L-M relation in][]{2002ApJ...567..716R,2011A&A...535A...4R}.
For the mass function bias estimates we used temperature scaling relations.
In the following study $\rm{M200c}$ is the mass of the cluster within the overdensity radius $r_\mathrm{200}$.

We investigated how the bias in the mass-proxy would affect the cluster mass-function expected for the eROSITA survey.
The mass function \citep[histogram of the number of clusters of a given mass, see][for a recent study]{2016MNRAS.456.2361B} of an observed cluster sample together with a selection function describing the sensitivity of the instrument for detecting certain cluster types can be converted into a real cluster population of the universe. This cluster mass function can be used to constrain cosmological parameters \citep[for a review on cluster cosmology, see e.g.][]{2011ARA&A..49..409A}.

The cluster mass $M_\mathrm{total}$ scales with temperature $T_\mathrm{X}$ as \citep[see e.g.][]{2001A&A...368..749F,2003MNRAS.342..163P,2009ApJ...692.1033V,2016A&A...592A...2P},

\begin{equation}
 M_\mathrm{total} \sim T_\mathrm{X}^{1.5}
\end{equation}

The bias of ${\frac{T_\mathrm{real}-T_\mathrm{obs}}{T_\mathrm{average}}=0.05}$ over a redshift range of $z = \mathrm{0.1-0.8}$, translates into an average mass bias of $\rm{7.5~per~cent}$.

Fig.\ref{fig:massfuntion} shows the systematic change in the cluster density in the redshift - mass plane assuming all masses were calculated from X-ray halo temperatures.

This systematic shift in the mass-function will cause a systematic offset in the derived cosmological parameters and has to be accounted for in cluster scaling relations for the eROSITA survey.

\begin{figure*}[ht!]
  \centering
  \resizebox{0.9\hsize}{!}{\includegraphics[angle=0,trim=0cm 0cm 0cm 0cm,clip=true]{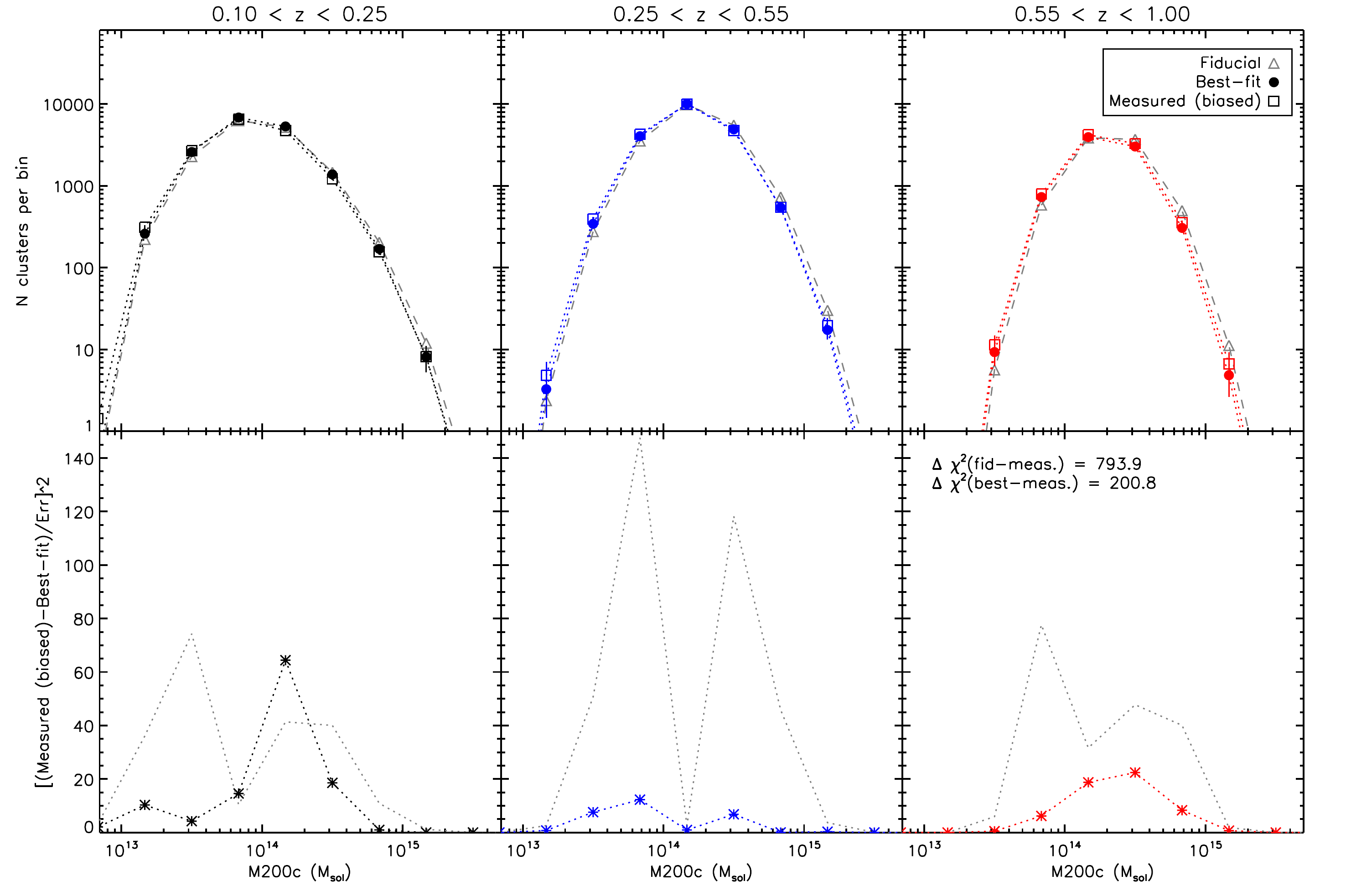}}
  \caption[eROSITA mass function bias in different redshift bins]{eROSITA mass function bias in different redshift bins. The top panels show the mass function (number of clusters in a given mass range $M_\mathrm{200c}$) in increasing redshift ranges (left: 0.1 to 0.25, middle: 0.25 to 0.55, right: 0.55 to 1.0). The triangles show the eROSITA mass function as expected in the standard cosmology \citep[fiducial cosmology WMAP9][]{2013ApJS..208...19H}. The open squares show the same mass function but with the estimated mass bias of about 7.5 per cent. The filled circles show the mass function for the best-fit cosmological parameters to the biased mass function. The errors on data-points are Poisson errors on the number of clusters per bin. The bottom panels show the offset significance between the fiducial model and biased mass function (grey dotted lines) and between the best-fit model and the biased mass function (crosses with coloured dotted line).}
  \label{fig:massf_bins}
\end{figure*}

To quantify the impact of the measured mass bias on the derived cosmological parameters the following technique was used. The overall mass function histograms of number of clusters in up to twelve mass bins was created for 3 different redshift ranges (see Fig.\ref{fig:massf_bins}). The offset significance between the fiducial and the biased mass-function is shown in the lower panels of Fig.\ref{fig:massf_bins}. This shows how the mass function changed due to the cluster mass bias. Covariance between $\rm{L_X}$ and $T_\mathrm{X}$ was neglected for creating the biased mass function.

The fiducial mass function was calculated using the \citet{2008ApJ...688..709T} mass function and the cosmological parameters from WMAP9 \citep[][]{2013ApJS..208...19H}. The fiducial cosmology parameters are $\rm{\Omega_M\approx0.28}$ and $\rm{\sigma_8\approx0.82}$. 
Using scaling relations the cluster masses can be converted to observables like X-ray temperature, flux, and source extent. Here we used relations found in the \emph{XMM-Newton} XXL-100 survey \citep[][]{2016A&A...592A...2P,2016A&A...592A...3G,2016A&A...592A...4L}. An instrument specific selection function describing detectability in flux and source extent can be used to relate the instrument specific observed mass function to the true mass function of clusters in the observed universe. 

The mass function can also be changed by varying the cosmological parameters $\rm{\Omega_M}$ and $\rm{\sigma_8}$. These parameters constrain the matter fraction of the total energy of the universe and the clustering amplitude of DM halos, respectively.

The number density of DM halos of different masses can be calculated for different cosmologies assuming purely gravitational collapse \citep[see early work by][]{1974ApJ...187..425P,1991ApJ...379..440B}. \citet{2008ApJ...688..709T} in recent work provided a universal function $\rm{f(\sigma)}$ describing the shape of a cosmological halo mass function mostly independent of redshift or cosmological model. 

As selection function the latest simulations of cluster detections was used in combination with latest scaling relation measurements from the \emph{XMM-Newton} XXL survey (Clerc et al., in prep). The function has decreasing detection probability at lower masses because low mass halos have lower X-ray luminosities and lower ICM temperatures.

\begin{figure}[ht]
  \centering
  \resizebox{0.99\hsize}{!}{\includegraphics[angle=0,trim=0cm 0cm 0cm 0cm,clip=true]{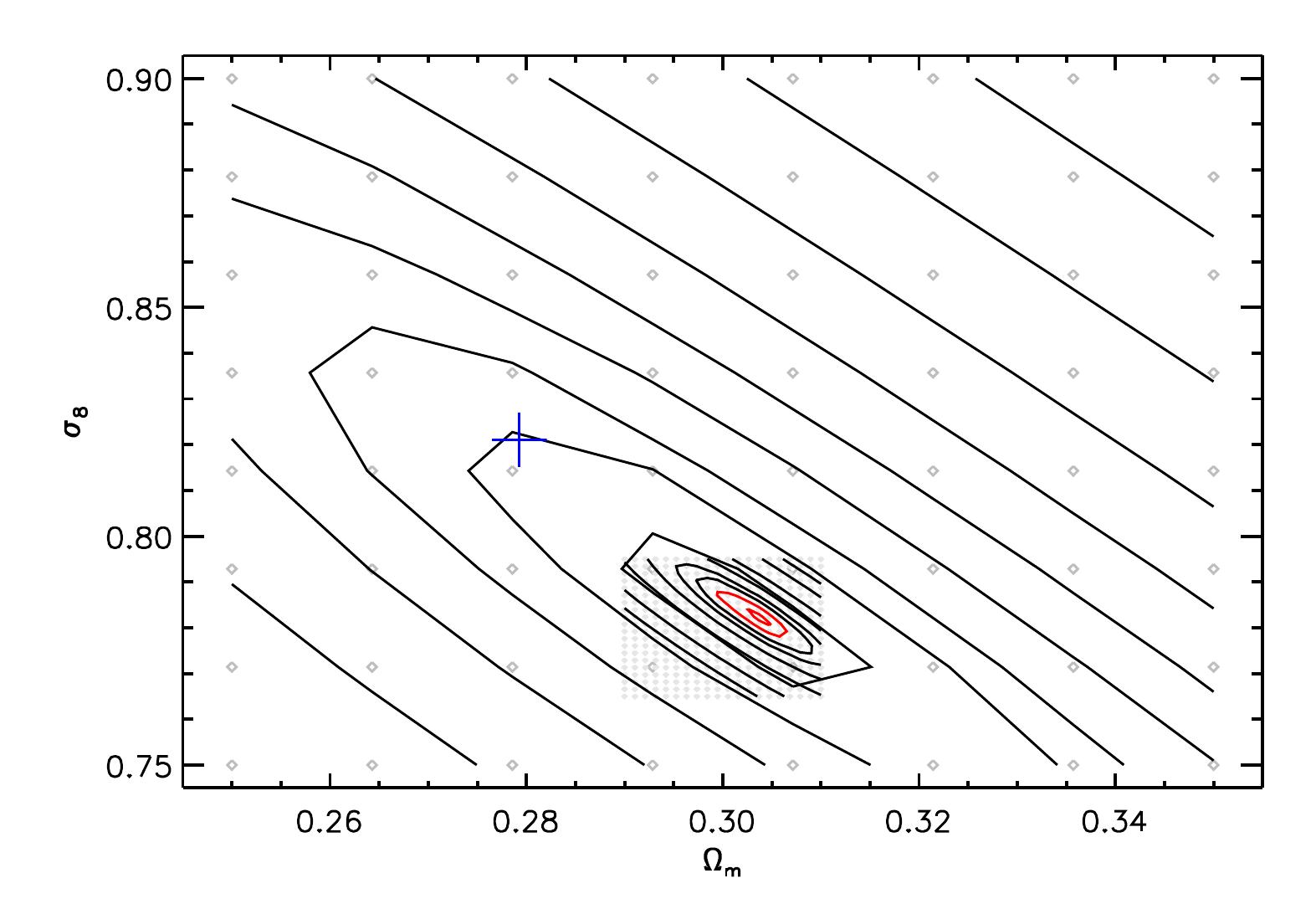}}
  \caption[eROSITA cosmology contours for biased mass function]{eROSITA cosmology contours for biased mass function for $\rm{\Omega_M}$ versus $\rm{\sigma_8}$. The cross shows the input (true) cosmology value. The black contours give the sum of the offset significance between the biased and fit mass function. The contours were obtained by interpolating the offset significance values calculated for the grid of cosmological parameters (see grey diamonds). The red contours show the 1 and 3 $\rm{\sigma}$ confidence level. The plot shows more coarse and finer contour levels around the best fit values.}
  \label{fig:cosmo_contours}
\end{figure}

For a grid of different $\rm{\Omega_M}$ and $\rm{\sigma_8}$ (see grey diamonds in Fig.\ref{fig:cosmo_contours}) an expected true mass function was created. For each combination of parameters the offset significance between true and biased mass-function was calculated to find the best fitting set of cosmological parameters \citep[for details on this method see][]{2012MNRAS.423.3561C}. Fig.\ref{fig:cosmo_contours} shows the best fit parameters and confidence contours around them ($\rm{\Omega_M}$ and $\rm{\sigma_8}$ were the only variable parameters of the fit).

The goodness-of-fit between the measured and simulated mass function was calculated as,

\begin{equation}
 \chi^2 = \sum \left\lgroup \frac{\rm{Measured}-\rm{Model}}{\rm{Error}} \right\rgroup^2
\end{equation}

which is the sum of the offset significance ($\rm{\chi^2}$ test) between the measured and model data points summed over all mass- and redshift-bins (see Fig.\ref{fig:massf_bins}). The annotated $\rm{\chi^2}$ values show that $\rm{\chi^2}$ between the measured (biased) mass function and the best fit simulated mass function compared to the fiducial mass function improved by $\mathrm{\Delta \chi^2 = 693.1}$. 

For an input cosmology with $\rm{\Omega_M=0.28}$ and $\rm{\sigma_8=0.82}$ we obtained a best fit of $\rm{\Omega_M\approx0.31}$ and $\rm{\sigma_8\approx0.78}$. This corresponds to a low bias of $\rm{\sim -5~per~cent}$ in $\rm{\sigma_8}$ and a high bias $\rm{\sim +10~per~cent}$ in $\rm{\Omega_M}$.

These trends can be understood because lower normalisation of the clustering $\rm{\sigma_8}$ would produce less massive halos consistent with the measurement bias.
If the measured masses are biased low this creates objects where the probability from the selection function is very low and thus the real expected number density at low masses is disproportionally boosted to increase the average expected matter density $\rm{\Omega_M}$.

Given that the error contours on the input cross in Fig.\ref{fig:cosmo_contours} would be similar to the ones for the biased value the offset is significant.

\section{Discussion}
\label{sec:discussion}

The most important mass proxy for galaxy cluster masses in X-ray surveys is the electron temperature of the ICM which can directly be measured from the X-ray spectra of a cluster. 
The ICM is generally assumed to be in thermal equilibrium in the observed regions and thus the electron temperature corresponds to the gas temperature. The temperature is obtained by fitting an emission model of a collisionally ionized plasma to the intrinsic X-ray spectrum of the source. Every X-ray instrument however has a slightly different response (i.e. detection efficiency) depending on incoming photon energy and position on the detector. 
The intrinsic source spectrum can be obtained by deconvolving the observed spectrum by the instrument response.

eROSITA is a rather soft X-ray telescope which means its effective collecting area is highest between $\rm{1-2~keV}$ and drops by a factor of about ten above 2\,keV. Because of this eROSITA is more sensitive to spectral features at low energies and thus more effective at detecting lower temperature gas. This biases the estimated average temperature when there is a second hotter gas component present in the spectrum \citep[as explained by][]{2013SSRv..177..195R}.

As many previous studies of deep X-ray observations of clusters have shown there is significant temperature structure in the ICM \citep[e.g.][]{2004ApJ...608..179R,2007MNRAS.381.1381S,2009MNRAS.399.1307M,2011A&A...528A..60L}.
Previous works studied the capability to measure the ICM temperature using different X-ray observatories.
\citet{2013SSRv..177..195R} reviewed the influence of multi-temperature ICM on the obtained average temperature of a cluster especially in the outskirts.
They found that eROSITA will significantly underestimate cluster temperatures by simulating a spectrum with two temperature components (0.5\,keV and 8.0\,keV). In a spectrum where the emission from cold and hot component was split 50/50 per cent the average temperature should be measured as 4.25\,keV. It was found however that in a single-temperature fit to the eROSITA spectrum the temperature was $\rm{\sim 1.5~keV}$ suggesting a low bias in temperature of about 60 per cent. The bias they measured is also varying with the assumed metallicity of the colder component. The temperature difference assumed in their simulations is more extreme than in the sample of clusters we analysed but demonstrates the expected trends for eROSITA. Our results show that in real clusters the bias towards lower temperatures due to substructures will be about 5 per cent.

\citet{2014A&A...567A..65B} tested how well temperature can be constrained for clusters in the eROSITA survey based on \texttt{XSPEC} \texttt{fakeit} simulations of isothermal clusters with a $\rm{\beta-model}$ surface brightness profile (including X-ray background).
The bias \citet{2014A&A...567A..65B} found does not include the substructure bias we investigated in this study. Their bias is mainly caused by the fitting method for the cluster X-ray spectra. In our study bias from the fitting method is cancelled out because we used the difference in temperature between two simulations with and without substructure.

\citet{2001ApJ...546..100M} found $\rm{\sim10-20~per~cent}$ lower temperatures measured in the 0.5 to 9.5\,keV band (similar to \emph{Chandra}) compared to emission weighted averages of hydrodynamic simulations. \citet{2005ApJ...618L...1R} found a bias of $\rm{\sim20-30~per~cent}$ lower temperatures from mock X-ray spectra compared to the emission-weighted average temperature from their simulations. leading to an underestimate of $\rm{\sigma_8}$ by $\rm{\sim10-20~per~cent}$ per cent. \citet{2006ApJ...640..710V} introduced an algorithm to better compare temperatures from simulations and from real cluster spectra.

This shows that the bias in comparison to high-resolution hydrodynamic simulations is higher than in our study but in the same direction.

\citet{2004MNRAS.354...10M} found temperatures measured from simulated \emph{Chandra} or \emph{XMM-Newton} X-ray spectra to be significantly lower than mass- and emission-weighted temperatures from their cluster simulations. The magnitude of the effect is similar to the one found in this study. Weighting the \emph{Chandra} map temperatures according to the spectral-like temperature (${T_\mathrm{sl}}$) \citep[][]{2004MNRAS.354...10M} instead of the average lowered the bias to $\mathrm{Bias}(T)\sim\mathrm{-2.5~per~cent}$. We found no significant temperature dependence in the offset between average- and spectral-like temperatures in this study.

The X-ray temperature is particularly important for cosmological studies because it is one of the best observable mass proxies for clusters of galaxies \citep[see e.g.][]{2001A&A...368..749F}. \citet{2003MNRAS.342..163P} have shown that the normalisation of the T-M scaling relation strongly influences the determination of the $\rm{\sigma_8}$ cosmological parameter.
They showed at high significance that a lower normalisation in the mass causes a lower predicted value for $\rm{\sigma_8}$. Lower normalisation of the T-M function is the same as a low mass bias due to temperature measurement bias as found in this study. 
They found that a ten per cent lower normalisation causes about a five per cent lower value for $\rm{\sigma_8}$ which is consistent with our results. Their results were obtained for fixed $\rm{\Omega_M=0.3}$.

An additional effect which has to be accounted for in the eROSITA mass calibration was presented by \citet{2016MNRAS.456.2361B}, who analysed the influence of baryons (mostly in form of hot gas in the ICM) on the mass function using data from the Magneticum simulations (Dolag et al., in preparation). Comparing the results of DM-only simulations and simulations including baryons they found that in case of eROSITA the change in the obtained mass-function could lead to an underestimate of about one per cent in $\rm{\Omega_M}$. 

\citet{2012MNRAS.422...44P} made the most detailed predictions for eROSITA cluster cosmology so far. They used the halo-mass function of \citet{2008ApJ...688..709T} obtained from N-body simulations of a $\Lambda CDM$ universe. Using L-M and T-M relations they converted the masses into detected photon count in eROSITA using early estimates of instrument properties. This allowed them to estimate uncertainties on cosmological parameters. They estimate that with eROSITA cluster counts, angular clustering measurements, photometric redshifts for all systems and cosmology priors from the \emph{Planck} mission it will be possible to obtain uncertainties of $\rm{\Delta\sigma_8 = 0.014 \lesssim 2~per~cent}$ and $\rm{\Delta\Omega_M = 0.0039 \lesssim 2~per~cent}$. 
At such high precision it will be crucial to correct for the bias we estimate to be 5-10 per cent especially when combining eROSITA with priors from other instruments.

The results demonstrated that a highly accurate mass calibration will have to account for differences in temperature bias among different cluster types. The bias is strongest for CC clusters and can be reduced by using core excised temperature measurements for the mass calibration in the eROSITA survey. Core excising will however only be possible for the brightest most extended subsample of clusters.

This study does not account for evolution in the cluster types with redshift. We assume the same sample of 33 clusters in the five redshift bins of our simulations. The eROSITA sample will contain much more low mass, and higher redshift clusters without strong CC. The simulations showed that the temperature bias due to substructure might be stronger in CC systems which are mostly evolved massive clusters. Therefore the overall bias in the eROSITA survey might be slightly lower. However since there was no evolution with redshift in our study and the trend for cluster types was not significant we assumed a constant bias throughout our sample and the future eROSITA sample. These assumptions cause some caveats (see also Sect. \ref{sec:bias_cav}) but the purpose of this work was to show a first order estimate of the cosmology impact that substructure can have in next stage cosmology studies.

The limited redshift range of our sample (plus the assumption of no evolution of the clusters between the simulated redshift bins) did not allow us to make predictions on dark energy constraints in the eROSITA survey.

\section{Summary and conclusions}
\label{sec:conclusion}

We isolated the bias effect that substructures in galaxy clusters have on temperature measurements and their impact on derived cosmological parameters.
This was achieved with eROSITA simulations of cluster spectra for a large sample of massive galaxy clusters in the four year all-sky survey. All simulations were based on real cluster observations with \emph{Chandra} and provide a representative sample of galaxy clusters in the local universe. 
In the redshift range of $\mathrm{0.1} \leq z \leq \mathrm{0.8}$ we measured a bias in the eROSITA cluster temperatures $T_\mathrm{X}$ of $\rm{-5.08\pm0.27}$ per cent and a bias in X-ray flux $F_\mathrm{X}$ of $\rm{-1.46\pm0.03}$ per cent.

Assuming temperature will be used as the primary eROSITA mass-proxy this causes a bias of about 7.5 per cent lower masses. This would cause an offset of $\rm{\sim -5}$ per cent in the cosmological parameter $\rm{\sigma_8}$ and $\rm{\sim +10}$ per cent in $\rm{\Omega_M}$ which the eROSITA cluster survey will be very sensitive to. This estimate was made assuming the cluster sample used in this study is representative over the covered redshift range.

Our findings emphasize that it will be crucial to calibrate cluster masses down to the per cent level to obtain confidence limits of $\rm{\sim2}$ per cent on the cosmological parameters investigated in this study. It will be equally important for precision dark energy studies with the eROSITA cluster sample.
The mass calibration accuracy will best be achieved using direct weak-lensing follow-up of eROSITA clusters after the first scan of the sky or by cross calibrating against mass measurements of other instruments like \emph{Chandra} or \emph{XMM-Newton}.

\begin{acknowledgements}

We thank the anonymous referee for very constructive comments that helped to improve the clarity of the paper. 
We thank J.J. Mohr for helpful discussions.
This research has made use of data obtained from the \emph{Chandra} Data Archive and the \emph{Chandra} Source Catalog, and software provided by the \emph{Chandra} X-ray Center (CXC) in the application packages CIAO, ChIPS, and Sherpa. 
This research has made use of NASA's Astrophysics Data System. 
This research has made use of the VizieR catalogue access tool, CDS, Strasbourg, France. 
This research has made use of SAOImage DS9, developed by Smithsonian Astrophysical Observatory. 
This research has made use of data and/or software provided by the High Energy Astrophysics Science Archive Research Center (HEASARC), which is a service of the Astrophysics Science Division at NASA/GSFC and the High Energy Astrophysics Division of the Smithsonian Astrophysical Observatory. 
This research has made use of the SIMBAD database, operated at CDS, Strasbourg, France. 

\end{acknowledgements}

\bibliographystyle{aa}
\bibliography{auto}

\end{document}